\documentclass[twocolumn,amsmath,epsfig,apsrev]{revtex4}
\usepackage{graphicx}
\usepackage{graphics}
\usepackage{dcolumn}
\usepackage{bm}
\usepackage{verbatim}
\usepackage{float}
\usepackage[usenames,dvips]{color}

\begin{document}

\title{Spin dynamics of molecular nanomagnets fully unraveled by four-dimensional inelastic neutron scattering}
\author{M. L. Baker$^{1,5}$, T. Guidi$^2$, S. Carretta$^3$, J. Ollivier$^1$, H. Mutka$^1$, H. U. G\"{u}del$^4$, G. A. Timco$^5$, E. J. L. McInnes$^5$, G. Amoretti$^3$, R. E. P. Winpenny$^5$ and P. Santini$^3$}

\affiliation{$^1$ Institut Laue-Langevin, BP 156, 6 rue Jules Horowitz, 38042 Grenoble Cedex 9, France.}
\affiliation{$^2$ ISIS facility, Rutherford Appleton Laboratory, OX11 0QX Didcot, United Kingdom.}
\affiliation{$^3$ Dipartimento di Fisica, Universit\`{a} di Parma, I-43124 Parma, Italy.}
\affiliation{$^4$ Department of Chemistry and Biochemistry, University of Bern, 3000 Bern, Switzerland.}
\affiliation{$^5$ School of Chemistry and Photon Institute, The University of Manchester, M13 9PL Manchester, United Kingdom.}
\date{June 28, 2012}

\begin{abstract}

Molecular nanomagnets are among the first examples of spin systems of finite size and have been test-beds for addressing a range of elusive but important phenomena in quantum dynamics. In fact, for short-enough timescales the spin wavefunctions evolve coherently according to the an appropriate cluster spin-Hamiltonian, whose structure can be tailored at the synthetic level to meet specific requirements. Unfortunately, to this point it has been impossible to determine the spin dynamics directly. If the molecule is sufficiently simple, the spin motion can be indirectly assessed by an approximate model Hamiltonian fitted to experimental measurements of various types.\\
Here we show that recently-developed instrumentation yields the four-dimensional inelastic-neutron scattering function $S({\bf Q},E)$ in vast portions of  reciprocal space and enables the spin dynamics to be determined with no need of any model Hamiltonian. We exploit the Cr$_8$ antiferromagnetic ring as a benchmark to demonstrate the potential of this new approach. For the first time we extract a model-free picture of the quantum dynamics of a molecular nanomagnet. This allows us, for example, to examine how a quantum fluctuation propagates along the ring and to directly test the degree of validity of the N\'{e}el-vector-tunneling description of the spin dynamics.

\end{abstract}

\maketitle
\noindent

Mesoscopic systems can exhibit typical quantum dynamical phenomena,
for instance by being able to tunnel through an energy barrier or by displaying long-lived coherent oscillations associated with superposition of states. This has attracted considerable interest for addressing fundamental issues and for the possible applications in quantum-information processing.
Molecular nanomagnets (MNMs) are spin clusters where the topology of magnetic interactions can be tailored precisely at the synthetic level.
They are metal-organic molecules containing a small number of magnetic ions whose spins are strongly coupled by exchange interactions. Shells of organic ligands provide magnetic separation between adjacent magnetic cores, which behave as identical and independent zero-dimensional units\cite{Gatteschi06}. The magnetic dynamics are characterized by strong quantum fluctuations and this makes MNMs of great interest in quantum magnetism as model systems to investigate a range of phenomena, such as quantum-tunnelling of the magnetization \cite{friedman96,thomas96,prokofiev98}, N\'{e}el-vector tunnelling (NVT)\cite{chiolero98,waldmannneel}, quantum information processing\cite{leuenberger01,QC1,simulazioni}, quantum entanglement \cite{wernsdorfer02,hill03,naturenano,candini10} or decoherence \cite{ardavan07,schlegel08,ardavan12}. Besides their fundamental interest, MNMs are also the focus of intense research for the potential technological applications as classical or quantum bits\cite{sessoli93,Gatteschi06,leuenberger01,QC1,simulazioni} and as magnetocaloric refrigerants\cite{evangelistiapl}.\\
A crucial aspect of the research on MNMs is the understanding of their low-temperature spin dynamics, especially of those aspects which are a direct manifestation of quantum mechanics like the tunneling of the N\'{e}el vector in antiferromagnetic rings. The most powerful technique to investigate the spin dynamics is inelastic neutron scattering (INS). INS measurements on MNMs are normally performed on polycrystalline samples. For these the cross-section is averaged over all molecular orientations, and only depends on the modulus of momentum transfer ($Q$). The resulting INS spectrum is composed of peaks whose energies yield the eigenvalues and whose areas contain information on the corresponding wavefunctions. The dynamics is then extrapolated by fitting experimental spectra to a suitable spin Hamiltonian $H$ and by performing calculations within the framework set by this model. However, this indirect approach does not work in many interesting cases when the Hilbert space dimension is too large to diagonalise $H$ or when a large number of model parameters makes the fit non-univocal. Even by using single-crystal samples, only limited and partially-integrated information on the ${\bf Q}$-dependence of the scattering function $S({\bf Q},\omega)$ can be obtained by traditional time-of-flight neutron spectrometers, i.e., carrying unitary detectors fitted on Debye-Scherrer rings.
The implementation of large arrays of position sensitive detectors in cold-neutron time of flight spectrometers\cite{ollivier,bewley}, together with the advances in software \cite{Horace}, has recently opened unprecedented possibilities in inelastic neutron scattering experiments on molecular nanomagnets allowing to determine the four-dimensional scattering function $S({\bf Q},\omega)$ in a vast portion of the reciprocal space. This provides a much more selective characterization of the MNMs when different candidate models can only be discriminated by the vectorial ${\bf Q}$-dependence of $S({\bf Q},\omega)$ \cite{waldmannmn12}.\\
\begin{figure}
	\centering
		\includegraphics[width=7 cm]{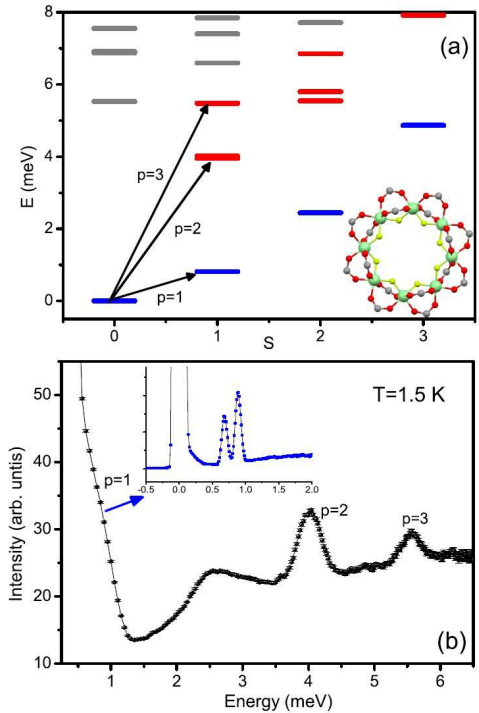}
	\caption{{\bf Magnetic energy spectrum of Cr$_8$ and zero-temperature inelastic neutron scattering transitions.} (a): Low-lying energy multiplets as a function of their total spin for the isotropic exchange Hamiltonian of Cr$_8$ (Eq. (\ref{eq:hamiltonian}) with $D=0$). Arrows indicate the three transitions seen by INS at zero temperature. All other transitions have negligible cross-section. Blue and red symbols indicate $L$- and $E$-band states respectively. The inset shows the core of Cr$_8$ (C$_{80}$Cr$_8$F$_8$D$_{144}$O$_{32}$ ; green=Cr, yellow=F, red=O, dark grey=C, D omitted). (b): measured low-$T$ INS spectra for Cr$_8$ with an incident neutron wavelength $\lambda = 3.1\AA$. The labels indicate the three peaks corresponding to the transitions reported in (a). The $p=1$ transition is partially hidden by the elastic signal. The inset reports higher-resolution measurements with $\lambda = 5 \AA$ showing the $p=1$ transition split by magnetic anisotropy.}
	\label{fig_bande}
\end{figure}
Here we show that the measurement of $S({\bf Q},\omega)$ over a large solid angle of wavevectors in high-quality single-crystals of MNMs permits us to directly determine the full pattern of real-space dynamical two-spin correlations without assuming any underlying model Hamiltonian. These correlations are the key quantities characterizing the magnetic dynamics. For instance, they determine the linear response of the system to an arbitrary magnetic field varying in space and time. We exploit Cr$_8$, a prototype antiferromagnetic (AF) ring (Fig. \ref{fig_bande} and Methods), as a benchmark to demonstrate the potential of this new approach for the  determination of the spin-dynamics in MNMs. The so-obtained correlation functions allow a model-free picture of the quantum dynamics of the ring. For example, the way a quantum fluctuation propagates along the ring or the degree of validity of a N\'{e}el vector tunneling description\cite{chiolero98,fe10} are determined.\\
\section{The C\lowercase{r}$_8$ prototype antiferromagnetic ring}
The Cr$_8$ molecule contains eight Cr$^{3+}$ ions ($s=3/2$) forming a nearly regular octagon \cite{Cr8VanSlageren,Cr8PRB}. It is one of the most studied MNMs both for its intrinsic interest and for being the precursor of a large family of heterometallic molecules and supramolecular complexes which are intensively studied for a number of fundamental and applicative issues \cite{Losshetero,chemcomm,rapidtorque,QC1,naturenano,candini10}. The spin-Hamiltonian of Cr$_8$ contains a dominating nearest-neighbor Heisenberg AF exchange and small axial anisotropic terms:
\begin{equation}\label{eq:hamiltonian}
H =  J\sum_{i=1}^8{\textbf{s}(i)\cdot\textbf{s}(i+1)}+D \sum_{i=1}^8 s_z^2(i)
\end{equation}
where $\textbf{s}(9)\equiv\textbf{s}(1)$, $J=1.46$ meV, $D=-0.038$ meV and the anisotropy axis $z$ is perpendicular to the ring plane\cite{Cr8PRB,Cr8PRL}.
The  ground state of Eq. (\ref{eq:hamiltonian}) is a nonmagnetic singlet (total spin $S=0$), and the low-lying excited levels are arranged into rotational bands (Fig.\ref{fig_bande}). The lowest one (called $L$-band) contains the ground state and approximately follows the Land\'e rule $E(S)=2JS(S+1)/N$\cite{Schnackrot,Gatteschi06}.
\begin{figure}
	\centering
		\includegraphics[width=9cm]{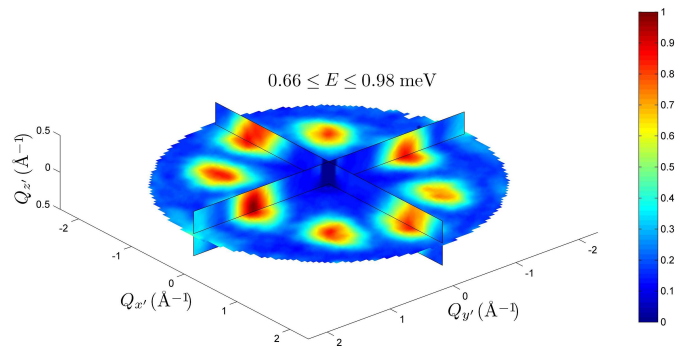}
	\caption{{\bf Constant energy plots of the neutron scattering intensity for the p=1 excitation.} Cuts are shown in the $Q_{x'}$-$Q_{y'}$, $Q_{z'}$-$Q_{y'}$ and $Q_{z'}$-$Q_{x'}$  planes, where the primed reference frame is defined in Fig. S1 (Supplementary information). The measurement was carried out with a $5 \AA$ incident neutron wavelength, and a sample temperature of 1.5 K.}
	\label{fig_3D}
\end{figure}
The second set of levels belongs to the so-called $E$-bands, which are also parabolic with respect to $S$ but shifted to higher energies.
\begin{figure*}
	\centering
		\includegraphics[width=18cm]{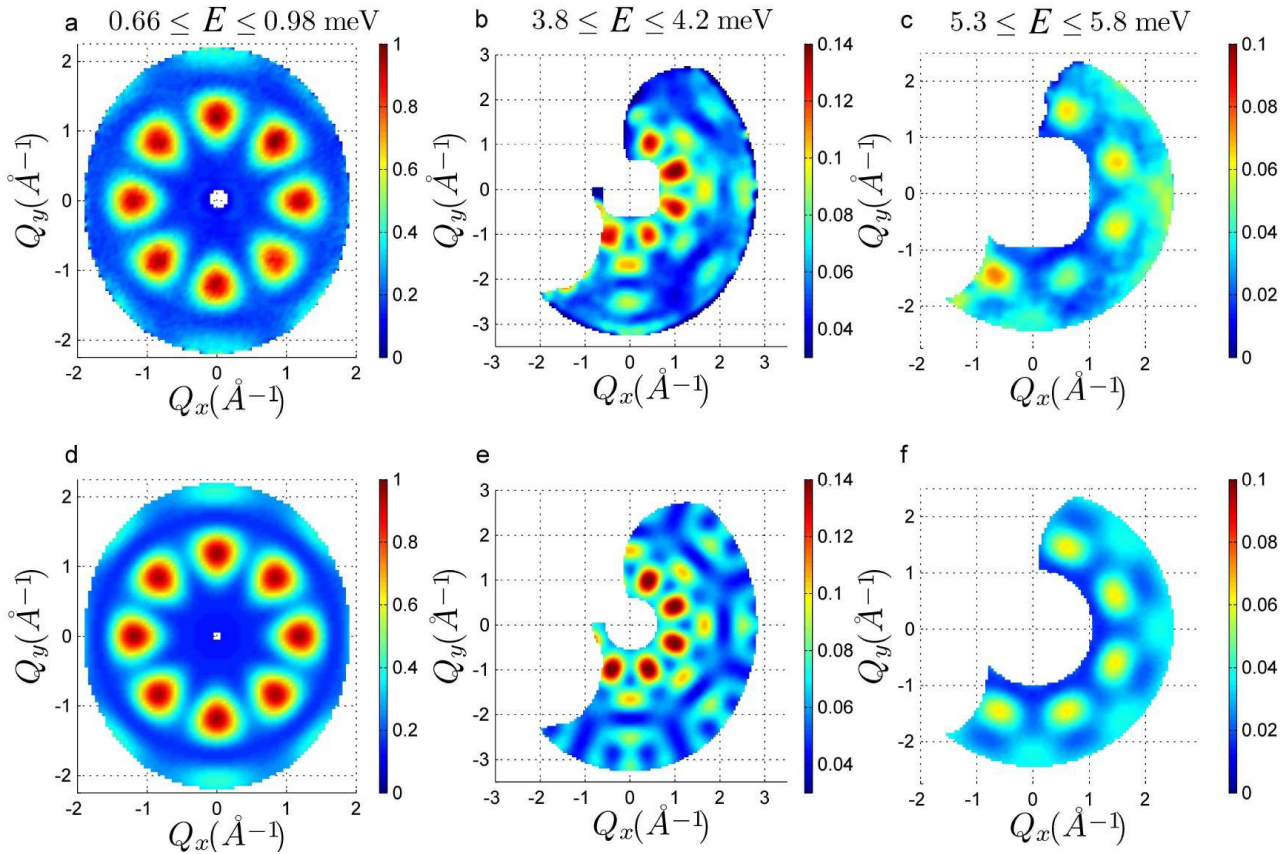}
	\caption{{\bf Constant energy plots of the neutron scattering intensity showing all the possible magnetic excitations in Cr$_8$ at 1.5 K.}  We show the dependency of intensity on two wavevector components $Q_x$-$Q_y$ lying in the ring's $x$-$y$ plane, integrated over the full $Q_z$ data range. a: Data from transition $p=1$, measured with a $5.0 \AA$ incident neutron wavelength. b-c, Data from transitions $p=2$ (b) and $p=3$ (c), measured with a $3.1 \AA$ incident neutron wavelength. d-f, fits to Eq. (\ref{eq:cross_section}) for excitations 1 (d), 2 (e) and 3 (f).}
	\label{fig_mappe_fit}
\end{figure*}
Due to their internal structure, $L$-band states can be excited by neutrons practically only to $L$- or $E$-band states\cite{Wald2001}. Hence at low temperature, where only the $S=0$ ground state is populated, three peaks are expected and observed in INS spectra\cite{Cr8PRB} (transitions marked by arrows in Fig. \ref{fig_bande}). Anisotropy produces small splittings of the otherwise degenerate $S$-multiplets and a tiny second-order mixing of different multiplets. For instance, the $L$-band $S=1$ triplet is split into an $|S=1,M=0\rangle$ singlet and an $|S=1,M=\pm 1\rangle$ doublet which are resolved by high-resolution INS measurements (inset of Fig. \ref{fig_bande}b). Splittings of the $E$-band triplets are smaller and not experimentally resolved.\\
\section{Inelastic neutron scattering and dynamical correlation functions}
The INS spectra in Fig. \ref{fig_bande}b are integrated over a wide range of momentum transfer ${\bf Q}$ and hence contain no direct information on the spatial structure of wavefunctions. This is usually indirectly inferred by fitting the spectra to a specific model Hamiltonian like Eq. \ref{eq:hamiltonian}.
By exploiting the new position-sensitive-detectors setup of the cold-neutron time of flight spectrometer IN5\cite{ollivier}, we have measured the detailed
${\bf Q}$-dependence of the scattering function $S({\bf Q},\omega)$ in a vast portion of reciprocal space.
\begin{figure*}
	\centering
		\includegraphics[width=15cm]{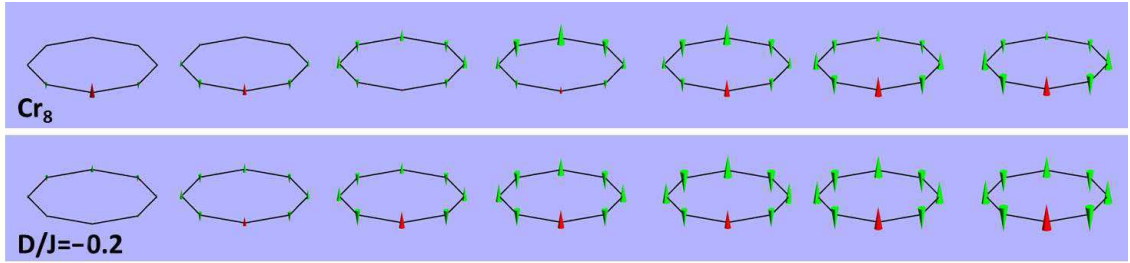}
\caption{{\bf Propagation of a local disturbance deduced from the present INS spectra and in a N\'{e}el-vector tunneling regime.} The frames show the time evolution of $\langle s_z(i)\rangle$ in an eight-spins ring after a delta-pulse perturbation $- b s_z(1)\delta(t)$ applied on the red ($d=1$) spin. Note that the ground state being a singlet, $\langle s_z(i)\rangle = 0$ just before $t=0$ (frame not shown).
Top : dynamics deduced from Eq. (5) using the present experimental Fourier coefficients (Table I) and frequencies. The delay between two frames is 1.9 10$^{-13}$ s, i.e., 1/32th of the longest oscillation period $2\pi/\omega_1$.
Bottom : results for a hypothetical ring in a Neel-vector tunneling regime, i.e., $D/J=-0.2$ corresponding to a tunnel action ${\cal S}_0\sim 7.6$. The delay between two frames is 1/32th of the tunneling oscillation period $2\pi\hbar/\Delta$, $\Delta$ being the tunneling gap.}
	\label{fig_film}
\end{figure*}
Fig. \ref{fig_3D} and the three upper panels of Fig.\ref{fig_mappe_fit} display examples of the ${\bf Q}$-dependence of $S({\bf Q},\omega)$ for the three low-temperature INS transitions. Since we would like to first address the effects of isotropic exchange, data for the $p=1$ transition are integrated over the energy range of the anisotropic splitting. Anisotropy will be addressed in the last part of the paper. Fig.\ref{fig_mappe_fit} shows that the variation of
$S({\bf Q},\omega)$ in the $Q_x$-$Q_y$ plane is characterized by several maxima and minima whose shape and positions depend on the specific transitions and
reflect the structure of the involved wavefunctions. More specifically, for a MNM with uniaxial anisotropy and $T\rightarrow 0$ \cite{MarshallLovesey}
\begin{eqnarray}\label{eq:cross_section}
&&S({\bf Q},\omega)\propto \sum_{\alpha=x,y,z} \left( 1-\frac{Q_\alpha^2}{Q^2}\right) \sum_p \sum_{d\ge d^\prime =1}^N F_d(Q)F_{d^\prime}(Q)\nonumber \\
&&\cos({\bf Q}\cdot {\bf R}_{d d^\prime}) \langle 0|s_\alpha(d)|p\rangle  \langle p|s_\alpha(d^\prime)|0\rangle\delta(\omega - E_p/\hbar)
\end{eqnarray}
where $F_d(Q)$ is the form factor for the $d$th ion\cite{booklet}, $|0\rangle$ and $|p\rangle$ are ground and excited eigenstates, respectively, $E_p$ are eigenenergies and ${\bf R}_{d d^\prime}$ are the relative positions of the $N$ magnetic ions within a molecule. As is well known, this formula can be rewritten in terms of $T=0$ dynamical correlation functions
\begin{equation}\label{eq:correl}
\langle s_\alpha(d,t) s_\alpha(d^\prime,0) \rangle = \sum_p  \langle 0|s_\alpha(d)|p\rangle  \langle p|s_\alpha(d^\prime)|0\rangle e^{-i \omega_p t},
\end{equation}
where $\omega_p =E_p/\hbar$. In fact the Fourier coefficients $c_{d d^\prime}^\alpha(\omega_p) = \langle 0|s_\alpha(d)|p\rangle  \langle p|s_\alpha(d^\prime)|0\rangle$ in Eq. (\ref{eq:correl}) coincide with the coefficients in Eq. (\ref{eq:cross_section}). While the values of $\omega_p$ are directly read out from the energies of the peaks in the INS spectrum, the $c_{d d^\prime}^\alpha(\omega_p)$ can be extracted from the data by fitting Eq. (\ref{eq:cross_section}) to the observed {\bf Q}-dependence of each peak. Indeed, for each value of $E_p$ these coefficients are the only unknown quantities in Eq. (\ref{eq:cross_section}).
In order to fix the scale-factor in Eq. \ref{eq:cross_section} we have exploited the sum rule
\begin{equation}
\langle 0|s^2(d)|0 \rangle = s(s+1) \equiv \sum_{\alpha,p}  c_{d d}^\alpha(\omega_p)
\end{equation}
with $s(s+1)=3.75$. \\
The Fourier coefficients $c_{d d^\prime}^\alpha(\omega_p)$ extracted by the fits of data integrated over the anisotropy splittings  are given in Table I. They do not depend significantly on $\alpha$ within experimental uncertainty, consistently with the first-order-perturbation effect of anisotropy in Cr$_8$.
\begin{table}
\caption{Fourier components of two-spin dynamical correlations $c_{1 d}^x(\omega_p)=c_{1 d}^y(\omega_p)=c_{1 d}^z(\omega_p)$ extracted from the present data (Exp column), and calculated by the isotropic Heisenberg model (Eq. (\ref{eq:hamiltonian}) with $D=0$).  The uncertainty on the experimental coefficients is of the order of 0.04 for $d>1$ and of 0.06 for $d=1$. Coefficients for $d=6,7,8$ are equivalent by symmetry to $d=4,3,2$, respectively.}
\begin{tabular}{c|c|c|c|c|c|c}
 & $p=1$ & & $p=2$ & & $p=3$ &    \\
\hline
$  $ & {\rm Exp} & {\rm Heis}  & {\rm Exp}& {\rm Heis} & {\rm Exp} & {\rm Heis}   \\
\hline
$ d=1 $ & 0.78 & 0.772  & 0.28 & 0.336 & 0.19 & 0.125  \\
\hline
$ d=2 $ & -0.77 & -0.772  &-0.15 & -0.173& 0.01 & 0.0  \\
\hline
$ d=3 $ &  0.78 & 0.772  & 0.0 & 0.0 &-0.14 &-0.125  \\
\hline
$ d=4 $ & -0.80 &-0.772  & 0.18 & 0.173 &-0.01 & 0.0  \\
\hline
$ d=5 $ &  0.81 & 0.772  &-0.32 &-0.336 & 0.1 & 0.125  \\
\end{tabular}
\label{table:av}
\end{table}
The three lower panels of Fig.\ref{fig_mappe_fit} show calculated ${\bf Q}$-dependencies corresponding to the best-fit Fourier coefficients $c_{1d}^\alpha(\omega_p)$.
The agreement with experimental maps is excellent.
The $d$-dependence of $c_{1 d}^\alpha(\omega_p)$, which determines the spatial pattern of correlations at the various frequencies, strongly varies with $\omega_p$. This is in line with the behavior expected from the model of Eq. \ref{eq:hamiltonian} (see Table I).\\
The information on the low-$T$ spin dynamics embedded in the dynamical correlation functions (Eq. (\ref{eq:correl})) can be visualized by exploiting the well-known link between these correlations and linear response functions\cite{MarshallLovesey}. The building blocks of such functions are the set of susceptibilities $\chi_{\alpha,d}^{\alpha',d'}(t)$, which provide the response of $s_\alpha(d)$ at time $t$ to a delta-pulse perturbation produced by a field $b$ of direction $\alpha'$ applied on spin $d'$ :
\begin{equation}\label{eq:susceptibility}
\delta H(t) = - b s_{\alpha'}(d^\prime)\delta(t) \rightarrow \langle s_\alpha(d)\rangle = b \chi_{\alpha,d}^{\alpha',d'}(t)
\end{equation}
with
\begin{eqnarray}\label{eq:linear_response}
\chi_{\alpha,d}^{\alpha',d'}(t)&=&\frac{i}{\hbar}\Theta(t)\langle\left[s_\alpha(d,t),s_{\alpha'}(d')\right]\rangle \nonumber \\
&\equiv&\delta_{\alpha ,\alpha'}\frac{2}{\hbar}\Theta(t)\sum_p c_{d d^\prime}^\alpha(\omega_p) \sin (\omega_p t),
\end{eqnarray}
$\Theta$ being the step function. The response here is diagonal ($\delta_{\alpha ,\alpha'}$) because of axial symmetry. As an example, we show in Fig. \ref{fig_film}  how a fluctuation on a given site at $t=0$, $\langle s_z(1)\rangle > 0$ (produced by a delta-pulse field on site 1) propagates in subsequent times along the ring.
The three-frequencies spectrum of Table I produces a wavelike motion of the magnetization, which propagates forth and back along the ring with a pattern determined by the precise values of Fourier coefficients. After the pulse is applied on site 1, waves propagating clockwise and counterclockwise interfere constructively at the opposite ($d=5$) site, thus producing a large value of local magnetization $\langle s_z(5)\rangle$ after about $6\times 10^{-13}$ s. Such constructive interference is lacking in the case of an odd-numbered AF ring. Here frustration would produce a complex time dependence when the two waves meet, which could be directly deduced by the present technique.\\
The measurement of the whole set of low-$T$ Fourier coefficients also allows us to extract equal-time correlation functions (i.e., Eq. (\ref{eq:correl}) with $t=0$), which are important quantities widely used to characterize the ground-state. The advantage here is that one can isolate the weak magnetic signal by integrating over the inelastic response. Such information would be extremely challenging to obtain directly by other techniques like diffuse scattering, even with polarization analysis. The staggered pattern of equal-time correlations (see Supplementary Information, Fig. S2) has an envelope slowly decreasing with distance as expected for the one-dimensional AF Heisenberg model with $s=3/2$, which has a quantum critical ground state with power-law correlations\cite{critical}.\\
\section{Anisotropy}
As discussed above, magnetic anisotropy in Cr$_8$ is small and its main effect is to cause a splitting of the $S=1$ triplet into a $|S=1, M=0 \rangle$ singlet and a $|S=1, M=\pm1 \rangle$ doublet. This splitting is witnessed by the presence of two separate peaks at 0.7 and 0.9 meV in the high-resolution INS spectrum (see the inset of Fig. \ref{fig_bande}b). We have investigated the effects of magnetic anisotropy on the spin dynamics by measuring the ${\bf Q}$-dependence of the neutron scattering intensity separately for the two low-energy peaks. The results are shown in the two upper panels of Figure S3 (see Supplementary information). The positions of the maxima are the same for both peaks but the relative intensities of the various maxima are different. In particular, the two $Q_x=0$ maxima are the most intense for the 0.7 meV transition but are the weakest for the 0.9 meV one. We have followed the same procedure as before to fit separately the ${\bf Q}$-dependencies of the two peaks. The so-extracted Fourier coefficients are reported in Table II and the corresponding calculated maps are displayed in the two lower panels of Figure S3. Table II shows that $c_{d d^\prime}^\alpha(\omega_p)$ are non-negligible only for $\alpha=z$ for the first peak and for $\alpha=x,y$ for the second peak. This is exactly the behavior expected for the axial Hamiltonian of Eq. (\ref{eq:hamiltonian}), and means that the low-frequency oscillations of $\langle s_z(d,t) s_z(d^\prime,0) \rangle$ occur at a somewhat smaller frequency than those of $\langle s_x(d,t) s_x(d^\prime,0) \rangle$ and $\langle s_y(d,t) s_y(d^\prime,0) \rangle$. We find a slightly larger weight of the $zz$ coefficients (first peak) with respect to the $xx$ and $yy$ ones (second peak), consistent with the asymmetry expected from the model of Eq. (\ref{eq:hamiltonian}) in presence of $S$-mixing\cite{Cr8PRB}.
\begin{table}
\caption{Fourier components of two-spin dynamical correlations $c_{1 d}^x(\omega_p)=c_{1 d}^y(\omega_p)$ and $c_{1 d}^z(\omega_p)$ directly extracted from the present data. The uncertainty on the experimental coefficients is of the order of 0.05. Calculations with the model of Eq. (\ref{eq:hamiltonian}) yield: alternating $\pm 0.86$ for the $zz$ coefficients and zero for the $xx$ and $yy$ ones (0.7 meV peak); alternating $\pm 0.73$ for the $xx$ and $yy$ coefficients and zero for the $zz$ ones (0.9 meV peak). The average of the $xx$, $yy$ and $zz$ coefficients is consistent with the values for the unresolved $p=1$ transition in Table I.}
\begin{tabular}{c|c|c|c|c}
 & $E=0.7$ meV & & $E=0.9$ meV     \\
\hline
$  $ & $xx$, $yy$ &  $zz$  & $xx$, $yy$ & $zz$    \\
\hline
$ d=1 $ &  0.00 & 0.82  & 0.76 & -0.02    \\
\hline
$ d=2 $ &  0.00 & -0.84  &-0.75 & 0.00  \\
\hline
$ d=3 $ &  0.00 & 0.81  & 0.72 & -0.02  \\
\hline
$ d=4 $ &  0.00 &-0.94  & -0.76 & 0.00  \\
\hline
$ d=5 $ &  0.01 & 0.95  &0.85 & -0.06   \\
\end{tabular}
\label{table:av}
\end{table}
\section{N\'{e}el-vector tunneling}
It has been proposed \cite{chiolero98} that the low-temperature dynamics of antiferromagnetic rings could be characterized by the coherent tunneling of the N\'{e}el vector ${\bf n}=\sum_d (-1)^d {\bf s}(d)$ between the $+\hat{z}$ and $-\hat{z}$ directions. This phenomenon is the AF counterpart of the tunneling of the magnetization (QTM) observed in single-molecule magnets \cite{Gatteschi06}. Interestingly, while for the latter the tunneling time is so long that the dynamics becomes overdamped due to dissipation into the nuclear-spins subsystem, the time scale of NVT in AF rings could be much shorter than the decoherence time. Within the semiclassical framework, the condition for the occurrence of N\'{e}el vector tunneling is that the action $S_0=Ns \sqrt{-2D/J}$ be much larger than unity. Hence, NVT is expected to occur in long AF rings with large anisotropy. While QTM can be probed by macroscopic magnetization measurements, NVT is elusive and a direct experimental demonstration of its occurrence is lacking. N\'{e}el vector tunneling has been recently shown to occur in a Fe$_{18}$ ring \cite{waldmannneel}, but only indirectly through powder INS and bulk thermodynamic measurements.\\
The possible occurrence of the NVT scenario in the low-$T$ dynamics of AF rings can be assessed by extracting dynamical correlations functions with the method presented in the paper. Indeed, in the case of NVT the two lowest eigenstates are close to the symmetric and antisymmetric superpositions of the two N\'{e}el states $|\uparrow \downarrow \uparrow ...\rangle$ and $|\downarrow \uparrow \downarrow...\rangle$. This implies that all the spins are rigidly locked into a two-sublattice pattern. If a local perturbing field is applied along $z$ on a single spin as discussed before, the expectation values of all the spins jointly increase in modulus (with an alternating pattern) with a practically instantaneous propagation of the perturbation. This behavior, shown in the lower panels of Fig. \ref{fig_film}, is completely different from the one we have experimentally found in Cr$_8$ (upper panels of Fig. \ref{fig_film}). Hence, these results demonstrate that the low-$T$ dynamics of Cr$_8$ is not characterized by NVT. Figure \ref{fig_film} shows that even for $|D/J|<1$ (but larger than in the case of Cr$_8$) NVT dynamics occurs. The present method could be applied to Fe$_{18}$ to deeply investigate its spin dynamics and directly show that it displays NVT, circumventing also the fact that the huge dimension of its Hilbert space prevents exact calculations.\\
\section{Conclusions and perspectives}

By exploiting the Cr$_8$ prototype AF ring as a benchmark, we have shown that the quantum spin dynamics of MNMs can be fully unraveled by four-dimensional inelastic neutron scattering. The possibility to extract all two-spin dynamical correlation functions opens remarkable perspectives in the understanding of crucial but elusive quantum phenomena in several classes of magnetic molecules. In particular, supramolecular complexes containing linked molecular nanomagnets have been demonstrated to display entanglement \cite{naturenano,candini10} and to be excellent candidates for implementing quantum-computation \cite{QC1} and quantum-simulation algorithms \cite{simulazioni}. The present method could be applied to reach a profound knowledge of the structure of the eigenstates of these complexes and of the entanglement between different molecular qubits.\\
Another emerging field is that of frustrated MNMs, whose magnetic correlations and spin dynamics are particularly complex. In fact, their energy spectrum has a structure with large degeneracies which cannot be rationalized in terms of simple paradigms (e.g., giant spins or rotational bands). However, a deep understanding of the spin dynamics of these systems is often hindered by the very large dimension of the associated Hilbert space. The approach described here can be exploited to determine the full pattern of dynamical spin correlations in some of these systems.\\

\noindent {\bf Acknowledgments}\\
M.L.B thanks the Japan society for the promotion of science for a postdoctoral fellowship. We acknowledge the Institute Laue-Langevin for funding and neutron instrument time. We thank the Institute Laue-Langevin technical staff, in particular R. Ammer who designed, machined and made alterations to the sample holders for the present experiments. TG thanks R. A. Ewings for the support with HORACE package.\\
This work was supported by the European Community through the ICT-FET Open Project MolSpinQIP, contract N. 211284,
and by Progetti di Interesse Nazionale (PRIN) project of the Italian Ministry of Research.\\
R. E. P. W. is supported by a Royal Society Wolfson Merit Award.\\ \\ \\

\noindent {\bf Methods}\\
The original procedure for preparation of compound  Cr$_8$F$_8$ [(O$_2$CC(CH$_3$)$_3$]$_{16}$[(CH$_3$)$_2$CO]$_2$ (1) reported in Gerbeleu, N. V. et al. Dokl. Akad. Nauk. SSSR 313, 1459, (1990), and then the modified version  published  in \cite{Cr8VanSlageren} produce the perdeuterated version of 1 in a very low yield (probably caused by the isotope effects on acidity of  deuterated  pivalic acid) and a new synthetic procedure for the deuterated compound {Cr$_8$F$_8$ [(O$_2$CC(CD$_3$)$_3$]$_{16}$} (2)was developed and is reported below.\\
Heating in a Teflon flask at 150 $^o$C for 14h while stirring of deuterated pivalic acid, ( (CD$_3$)$_3$CCO$_2$H , 11.0 g, 98.94 mmol), hexamethylenetetramine ( C$_6$H$_{11}$N$_4$, 0.6 g, 4.28 mmol) and chromium(III) fluoride tetrahydrate (CrF$_3$ .4H$_2$O, 3.0 g, 16.57 mmol) resulted  in a green microcrystalline product. After this, the flask was cooled to room temperature, and acetonitrile (20ml ) was added to complete the precipitation. The product was collected by filtration, washed with acetone, dried, and then was extracted with hexane (50 mL), and finally purified on a silica gel column using toluene as the eluent. The first fraction from the column was collected and the solution was evaporated to dryness. Yield: 3.15 g (65 \%). Elemental analysis calculated (\%) for C$_{80}$Cr$_8$F$_8$D$_{144}$O$_{32}$ : Cr 17.85, C 41.22 ; found: Cr 17.80,  C 40.99. ES-MS (sample dissolved in THF; spectra run in CH$_3$OH) : 2352$^+$  [M+Na]$^+$ ,  2384$^+$ [M+CH3OH+Na]$^+$.\\
Large crystals of  Cr$_8$ were obtained by very slow evaporation of  hexane  solution of 2 at ambient condition. The crystal structure is tetragonal (space group $P4212$) with $a=b=20.093(2)\AA $ and $c=16.801(2)\AA$.\\

INS experiments were performed on the IN5 time-of-flight inelastic neutron spectrometer\cite{ollivier} at the high-flux reactor of the Institute Laue-Langevin. The IN5 instrument has a 30m$^2$ position sensitive detector
divided in 10$^5$ pixels, covering 147 degrees of azimuthal angle and $\pm 20$
degrees out-of plane. A 0.24 g Cr$_8$ single crystal was aligned as in Fig. S1 (Supplementary Information) and  measurements were taken by rotating the crystal (in steps of 1 degree) about the vertical z' axis.
Incident neutron wavelength of 5.0 and $3.1 \AA$ were selected to probe measured excitations with energy resolutions, at zero energy transfers, of 0.079 and 0.38 meV respectively.
The uniformity of the detector sensitivity was ensured by measurement of
a standard vanadium sample.\\
Measurements for different rotations angles were combined using the HORACE analysis suite\cite{Horace}. $S({\bf Q}^\prime,\omega)$ was rotated by 37.8 degrees about the $Q_{y^\prime}$ axis to analyze the data in the $(x,y,z)$ molecular reference frame. The two molecules in the unit cell (Fig. S1) were included in the calculation of the scattering function and fitting of Eq.(\ref{eq:cross_section}) to the {\bf Q} dependence of the observed transitions was performed using HORACE. Some compensation for inhomogeneity in the scattering intensity due to self-shielding is performed, based on the variation of incoherent scatting intensity with sample orientation. \\  \\

\noindent {\bf Author contributions}\\
M.L.B., T.G., S.C., J. O., H.M., H.U.G. and G.A. did the experiment on a crystal synthesized by G.A.T. after discussion with E.J.L.M. and R.E.P.W.
Data treatment was made by M.L.B., T.G., J. O., H.M., and data simulations and fits were performed by M.L.B., T.G. and S.C.\\
S.C., G.A. and P.S. developed the idea to use four-dimensional INS measurements for extracting dynamical correlation functions of molecular nanomagnets. S.C., G.A. and P.S. also did theoretical calculations and wrote the manuscript with input from all coauthors.

\newpage
\section{Supplementary Figures}

\setcounter{figure}{0}

\begin{figure}[H]
	\centering
		\includegraphics[width=6cm]{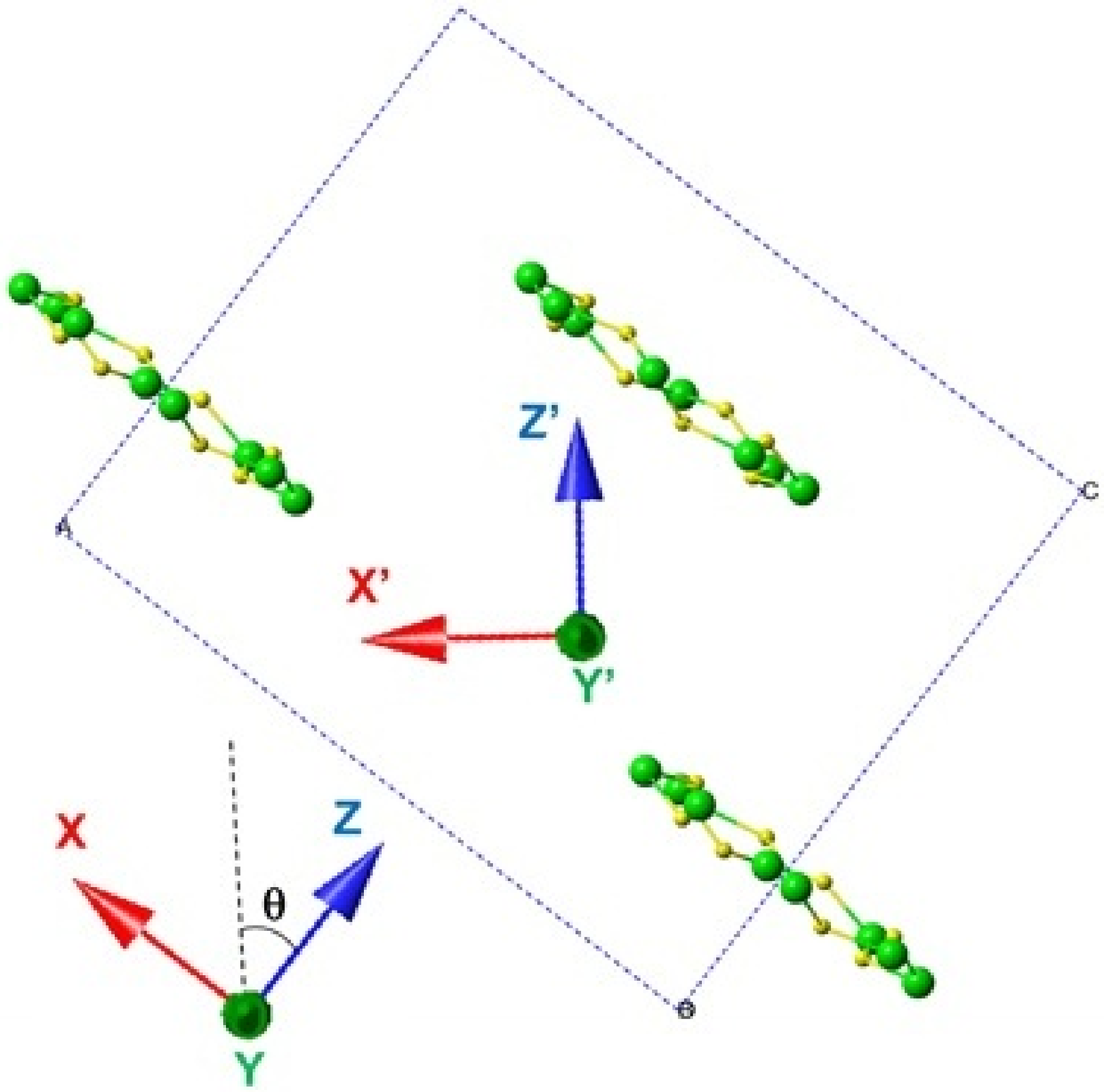}
	\caption{Orientation of the Cr$_8$ single crystal for the IN5 experiment. $(x',y')$ defines the  laboratory horizontal scattering plane and $(x,y)$  is the plane parallel to the ring. The dotted line is  the unit cell. The crystal was mounted with the [1 0 1] direction  approximately perpendicular to the scattering plane. The angle between z, perpendicular to the ring plane, and the laboratory z' direction is 37.8 degrees. }
\end{figure}

\begin{figure}[H]
	\centering
		\includegraphics[width=5cm]{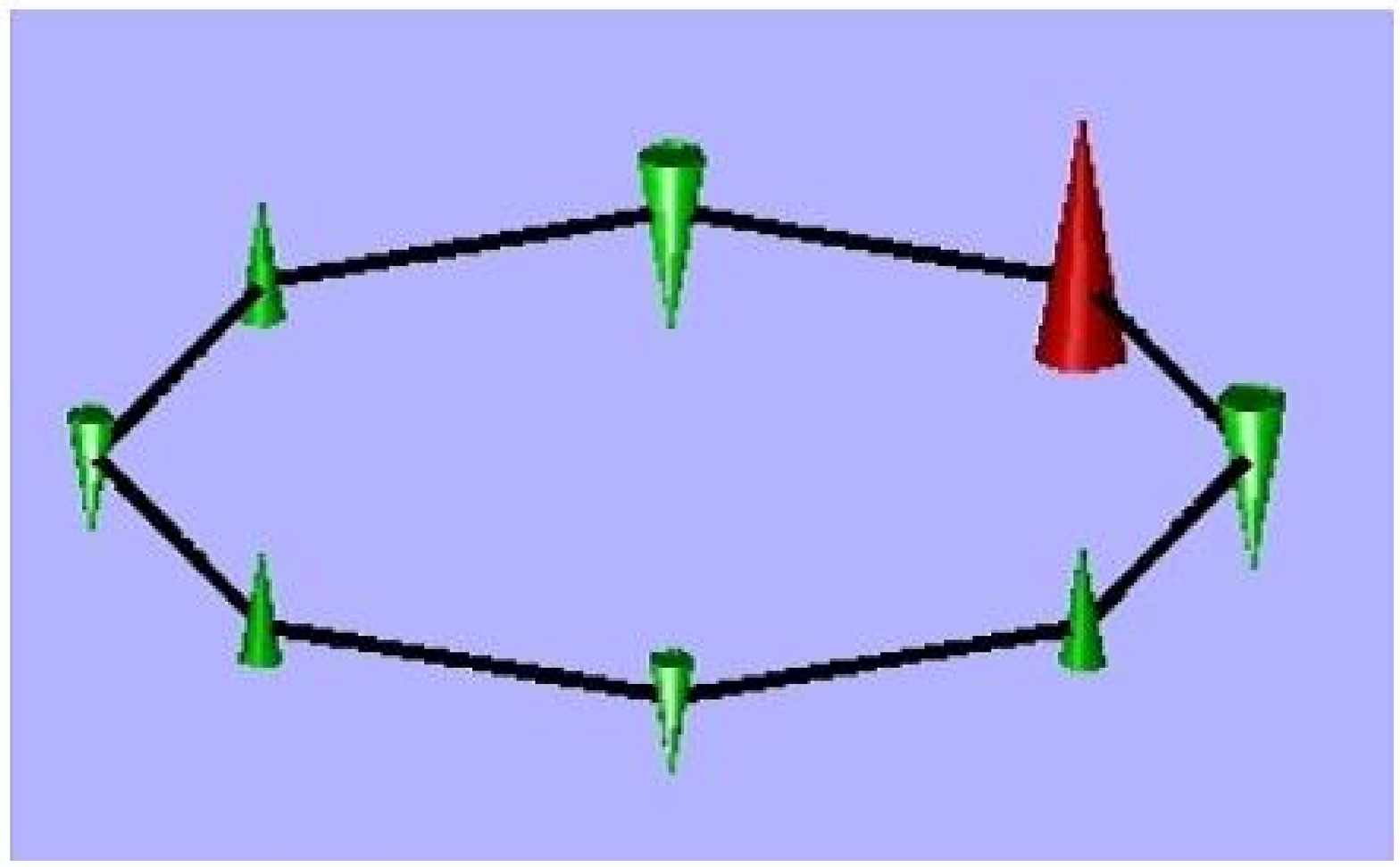}
	\caption{Illustration of the equal-time correlations $\langle s_z(d)s_z(1)\rangle$ deduced from Table I of the main text and mapped onto the length of the cones. The red cone corresponds to $d = 1$.}
\end{figure}

\begin{figure}[H]
	\centering
		\includegraphics[width=9cm]{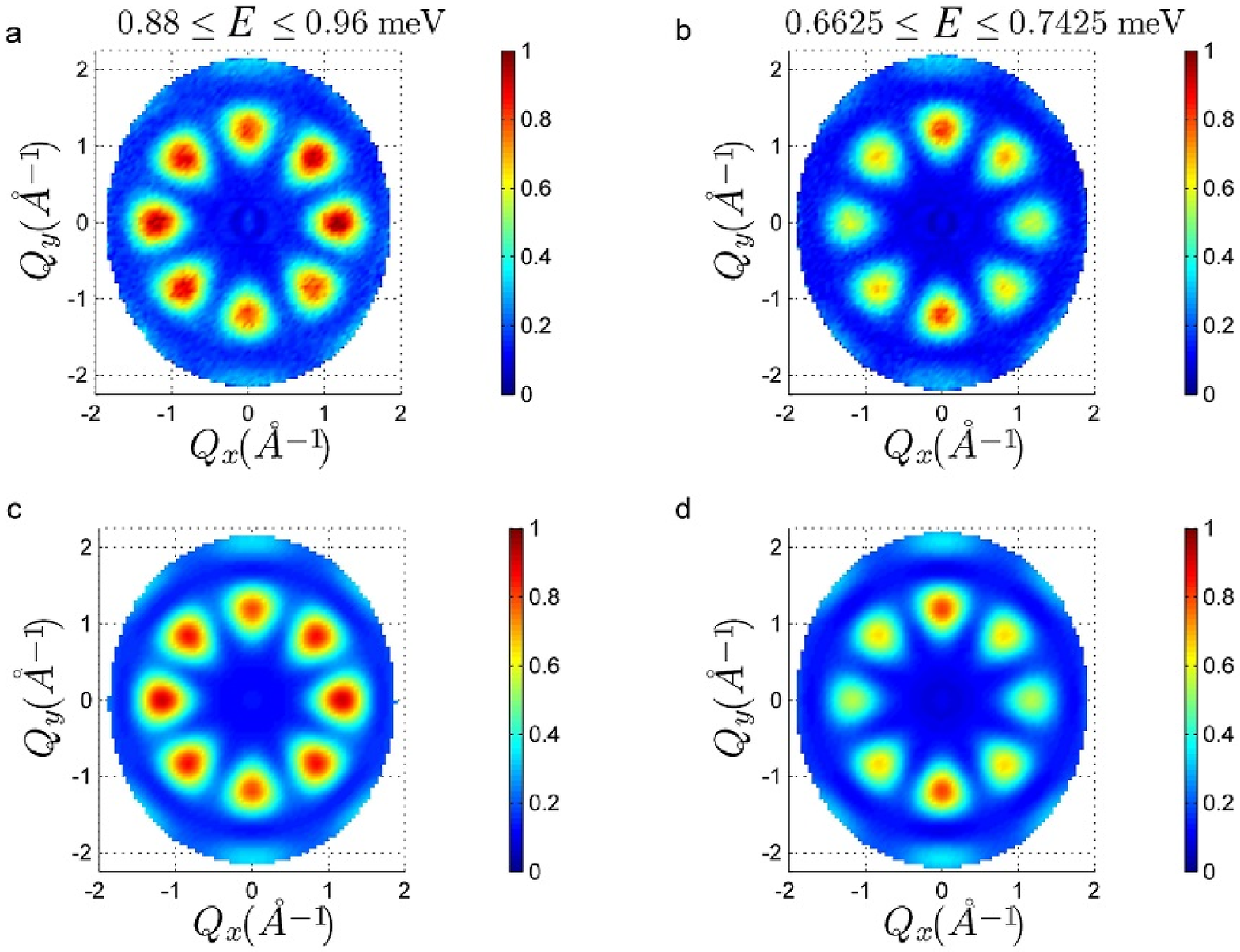}

	\caption{Constant energy plots of the neutron scattering intensity for the two resolved excitations in the p=1 peak (see the inset in Fig. 1 of the main text). As in Fig. 3 of the main text, we show the dependency of intensity on two wavevector components $Q_x-Q_y$ lying in the ring's $x-y$ plane, integrated over the full $Q_z$ data range. a (b): Data from the higher-E (lower-E) subpeak measured with a 5 \AA incident neutron wavelength. The positions of the maxima are the same for both peaks but the relative intensities of the various maxima are different. In particular, the two $Q_x = 0$ maxima are the most intense for the 0.7 meV transition but are the weakest for the 0.9 meV one.
c-d: fits to Eq. 2 (coefficients are reported in Table II in the main text).}
\end{figure}

%
%

\end{document}